# Soft phonons and structural phase transition in superconducting $Ba_{0.59}K_{0.41}BiO_3$

H. J. Kang[1,2,3*], Y. S. Lee[4], J. W. Lynn[1], S.V. Shiryaev[5] and S. N. Barilo[5]

[1]NIST Center for Neutron Research, National Institute of Standards and Technology, Gaithersburg, MD 20899, USA

[2]Department of Materials Science and Engineering, University of Maryland, College Park, MD 20742, USA

[3]Department of Physics and Astronomy, Clemson University, SC 29634

[4]Department of Physics, and Center for Materials Science and Engineering, Massachusetts Institute of Technology, Cambridge, Massachusetts 02139, USA

[5]Institute of Solid State and Semiconductor Physics, Academy of Science, Minsk 220072, Belarus

*Email: hkang70@clemson.edu

We have observed a softening of phonons and a structural phase transition in a superconducting $Ba_{0.59}K_{0.41}BiO_3$ ($T_c$ = 31 K) single crystal using elastic and inelastic neutron scattering measurements. The soft phonon occurs for the [111] transverse acoustic mode at the zone boundary. The phonon energies in this vicinity are found to continuously decrease with decreasing temperature from above room temperature to 200 K, where a structural phase transition from cubic to tetragonal symmetry occurs. The overall results are consistent with previous data that reported phonon softening and a (0.5, 0.5, 0.5) type superstructure in several $Ba_{1-x}K_xBiO_3$ systems. However, we also find weak (0.5, 0.5, 0) type superstructure peaks that reveal an additional component to the modulation. No significant change related to the superconductivity was observed for the soft phonon energies or linewidths.

PACS: 74.70.Dd, 74.25.Kc, 78.70.Nx, 61.05.F-

## I. Introduction

$Ba_{1-x}K_xBiO_3$ (BKBO) is a very interesting superconductor system to study, as it possesses one of the highest superconducting transition temperatures ($T_c$ = 31 K) for a conventional phonon-mediated superconductor, only being surpassed by $MgB_2$ and $Cs_xRb_yC_{60}$.[1,2] Higher $T_c$ values have been observed in the copper oxide (cuprate) and iron-based superconductors,[3-10] which are planar materials where magnetic fluctuations are believed to play an essential role in Cooper pair formation. In contrast, BKBO has the basic cubic perovskite crystal structure,[11,12] and does not exhibit any kind of cooperative magnetic behavior.[13] Consequently, electron-phonon coupling is the expected mechanism of superconductivity, and indeed strong evidence of electron-phonon coupling has been found in several experiments, including a large oxygen isotope effect[14] and the softening of phonons in the BKBO system.[15] Inelastic neutron scattering measurements by Braden *et al.* show that the oxygen optical phonon along the [100] direction shifts to much lower energy near the middle of the zone in BKBO (x = 0.4), which is the composition where the highest $T_c$ is realized.[15] In the present investigation we were hoping to find a dramatic 'resonance' effect in the phonon excitation spectrum that is directly associated with the formation of the superconducting phase, as has been found in the electron-phonon superconductors $YNi_2B_2C$,[16] and $LuNi_2B_2C$,[17] and in analogy with the magnetic resonance found in the cuprates [18-25] and iron-pnictides.[26,27] It was thought that the dramatic changes in the phonon lifetime were associated with a Fermi surface nesting, but it turns out that it is a more general wave vector phenomenon originating from the electron-phonon interaction near the superconducting gap 2Δ,[28] and is in quite good quantitative agreement with the theory of Allen *et al.*[29] Measurements of the phonon energies and linewidths are the subject of a separate study; here we report some new features of the structural phase transition.

One of the signatures of phonon mediated superconductivity is the change of acoustic phonon lifetimes and energies around the superconducting transition temperature.[30] For conventional Bardeen-Cooper-Schrieffer (BCS) superconductors, phonons with energies lower than the superconducting gap energy 2Δ should shift to a lower energy in the superconducting state. In addition, the lifetime of these phonons should increase, since phonons with energy less than $2\Delta$ are not able to break Cooper pairs and hence cannot be absorbed by superconducting electrons.[30] The increase of phonon lifetimes is exhibited as a narrowing of the energy width of a phonon peak observed in neutron or x-ray inelastic scattering. Above the gap, on the other hand, a decrease in the lifetime is observed due to the 'piling-up' of the electron density of states in the superconducting state.

The shift of phonon energies and a narrowing of the peak width when the system was cooled below $T_c$ indeed has been observed by neutron scattering in a number of conventional electron-phonon superconductors such as $Nb_3Sn$ and Nb.[31, 32] Similar superconductivity induced phonon softening has also been observed in the unconventional high temperature cuprate $YBa_2Cu_3O_7$.[33, 34] This phonon softening occurs at an energy close to the superconducting gap energy $2\Delta$. Soft phonons have also been observed in the rare earth nickel boride carbides $LuNi_2B_2C$ [35] and $YNi_2B_2C$ [16] compounds, which are nonmagnetic systems with a relatively high $T_c$. In the $LuNi_2B_2C$ compound, a substantial decrease of the acoustic and optical phonon energies was observed around the zone boundary for the [100] transverse mode. A more detailed temperature dependence was observed in the $YNi_2B_2C$ compound by Kawano *et al.*, who showed that the softening of the optical phonon [100] transverse mode near the zone boundary continues until the temperature gets to $\approx 30$ K.[16] The peak position of this soft phonon centered at E = 7 meV shows no further change below 30 K. Interestingly, a dramatic change in the lifetime causes a sharp 'resonance-type' scattering peak to develop for an energy close to $2\Delta$ (4.3 meV). As the temperature traversed $T_c$, its intensity followed a BCS-type order-parameter-like curve below $T_c$.[16] In addition, as the intensity of this peak increased the intensity of the higher-energy optical phonon became weaker, suggesting a direct connection between the two. They further showed that this scattering was directly related to superconductivity by measuring the field dependence, where the intensity monotonically decreased with increasing field and could be extinguished completely when the magnetic field exceeded the superconducting upper critical magnetic field. It will be interesting to investigate similar phonon softening and lifetime effects in the high temperature superconducting BKBO system.

## II. Experimental details

The $Ba_{0.59}K_{0.41}BiO_3$ single crystal was grown using a modified method of electrochemical deposition (seeded growth method).[36] The crystal was 2 cc in volume and in the shape of a flat plate, with a relatively narrow bulk superconducting transition temperature $T_c$ = 31 K.[37] The crystal structure is cubic (space group: *Pm-3m*) with a lattice parameter a = 4.28 Å at room temperature.

We performed inelastic neutron scattering measurements to determine the phonon dispersion curves of our superconducting BKBO single crystal with the triple axis spectrometers BT2 and BT9 at the NIST Center for Neutron Research. Pyrolytic graphite (PG(002)) crystals were employed as a monochromator and analyzer, with a final energy fixed at $E_f$ = 14.7 meV. Söller collimations of 40'-48'-40'-120' full-width-at-half-maximum (FWHM) were used, and a PG filter was placed after the sample to remove higher order wavelength contaminations. The crystal was aligned in the (HHL) scattering plane defined by scattering vectors [110] and [001] to measure phonon dispersion curves along high symmetry directions of [100], [110], and [111]. The momentum transfer **Q** = (H, K, L) is labeled in reciprocal lattice unit (r.l.u.), where r.l.u. = a* = 2π / a. The sample was sealed in an aluminum can with helium gas for heat exchange and its temperature was controlled using a closed cycle refrigerator.

## III.   Results

We measured low lying phonon dispersion curves along three high symmetry directions. A measurable change with temperature for the transverse acoustic [111] phonon branch was observed close to the zone boundary. Figure 1 shows selected energy scans at positions close to the zone boundary as a function of temperature. The data in Fig. 1 (a) indicate that the phonon peak close to the zone boundary at **Q** = (0.53, 0.53, 1.53) continuously shifts its position to lower energy when the temperature is decreased from 315 K to 200 K. Below 200 K, there is no further change in the position of the peak. Energy scans close to the zone boundary at $\mathbf{Q} = (0.55, 0.55, 1.55)$ (Fig. 1 (b)) clearly show that the peak position does not change from 140 K down to 8 K. Note in particular that there is no observable change across the $T_c = 30$ K. This suggests that this phonon softening is not directly related to the superconducting pairing, but instead is associated with a structural transition from cubic to tetragonal symmetry, which was reported by Braden et al.[38] They observed (0.5, 0.5, 0.5) type superstructure peaks (in cubic notation) from their x-ray and neutron diffraction measurements. The (0.5, 0.5, 0.5) type superstructure peak can arise from either doubling of the unit cell in all three direction ($2a_c \times 2a_c \times 2a_c$ cubic structure), or a $\sqrt{2}a_c \times \sqrt{2}a_c \times 2a_c$ tetragonal structure, where $a_c$ is the cubic lattice constant. The peak splitting observed by Braden et al. from their x-ray powder diffraction data leads to a $\sqrt{2}a_c \times \sqrt{2}a_c \times 2a_c$ tetragonal structure (space group: *I4/mcm*) to describe the new superstructure. The room temperature cubic structure of BKBO is shown in Fig. 2 (b), where we see that each Bi atom is surrounded by 6 oxygen atoms, forming $BiO_6$ octahedra. Braden et al. suggest that the structural transition is characterized by a $BiO_6$ octahedra rotation along the cubic [001] direction as shown in Fig. 2 (d).[38] The small (black) square is the basal plane for a cubic unit cell and the large (blue) square is the basal plane for a tetragonal unit cell with lattice parameters, $a_t = \sqrt{2}a_c$ and $c_t = 2a_c$. The tetragonal structure suggested by Braden et al. is shown in Fig. 2 (c).

We have observed the (0.5, 0.5, 0.5) type superstructure peaks in our elastic neutron measurements, in agreement with Braden, et al. The [HHH] scan around the (0.5, 0.5, 2.5) position and its integrated intensity as a function of temperature are shown in Fig. 3 (a) and (b). This superstructure peak develops when the temperature is decreased from above room temperature and its intensity increases with decreasing temperature. The integrated intensity of this peak shows that the structural transition is continuous with an ordering temperature $T_s \approx 200$ K. It is also apparent that there is some distribution of transition temperatures for this large crystal. The energy scan around the elastic position of the (0.5, 0.5, 1.5) peak and its integrated intensity in Fig. 3 (c) and (d) show the same result. In addition, we also found a new set of superstructure peaks located at (0.5, 0.5, 0) type positions. The (1.5, 1.5, 1) peak (Fig. 2 (a)) develops as the temperature is lowered below 200 K, and is much weaker than the primary (1/2,1/2,1/2) type reflections. Systematic checks as a function of wavelength indicated that these new peaks are not due to multiple Bragg scattering. This (0.5, 0.5, 0) type superstructure peak is not allowed in the *I4/mcm* space group suggested by Braden *et al.*, where the relation of Miller indices between tetragonal and cubic is $(H_t, K_t, L_t) = (H_c - K_c, H_c + K_c, 2L_c)$. Then the (0.5, 0.5, 0) type peak in cubic notation becomes the (0, 1, 0) type peak in tetragonal notation, which is not allowed in a body centered tetragonal structure where allowed peaks should satisfy the condition H + K + L = even. This may indicate that the octahedral rotation around [001] is not precisely a rigid rotation, but has some additional distortion which breaks the high symmetry. One of the subgroups of *Pm-3m* that allows both the (0.5, 0.5, 0.5) and (0.5, 0.5, 0) type peaks is *P4/mmm* (a primitive tetragonal structure).

Figure 4 shows the transverse acoustic phonon dispersion curve along the [111] direction as a function of temperature. Most of the softening occurs close to the zone boundary when the temperature is decreased. We have shown that the softening of the phonons stops around the structural transition temperature $T_s \approx 200$ K, well above $T_c$, and there is no change across $T_c$. The phonon softening may indicate that there is a strong electron-phonon interaction in this regime of k-space and this interaction causes the structural phase transition in BKBO system. This behavior is typical of strongly coupled conventional superconductors that have relatively high $T_c$,[35, 39] but having no significant change of these phonon modes across $T_c$ suggests that the softening of transverse acoustic phonons observed along [111] direction is not directly related to superconductivity. We note that the strong electron-phonon interaction in BKBO system is also responsible for a pseudogap that was observed by high resolution angle-integrated photoemission spectroscopy.[40] The pseudogap was closed at 300 K and 150 K for x = 0.33 and 0.46, respectively.

## IV. Discussion and Summary

The anomaly observed in optical and acoustic phonons in the BKBO system is a very promising signal for the role of phonons in superconductivity. The phonons are surely the origin of the pairing, so changes of the phonon energies and/or lifetimes are expected across $T_c$. Therefore, further detailed investigations of the lattice vibrations are warranted and underway. For the present neutron measurements in the (HHL) scattering plane, we have measured all the longitudinal and transverse acoustic phonons propagating along three high symmetry directions: [100], [110], and [111], with the exception of the [110] transverse acoustic mode with ionic displacements along the [1-10] direction (TA1), which requires a different scattering plane. A soft phonon occurs for the [111] transverse acoustic mode at the zone boundary, with the phonon energy continuously decreasing with decreasing temperature from above room temperature to 200 K, where a structural phase transition from cubic to tetragonal symmetry occurs. The overall results are consistent with previous data that reported phonon softening and a (0.5, 0.5, 0.5) type superstructure for several $Ba_{1-x}K_xBiO_3$ compositions near the optimal x, but with a small additional distortion characterized by (0.5, 0.5, 0) type superstructure peaks.

**Acknowledgments**

The work at Clemson University is supported by DOE/EPSCoR under Grant No. DE-FG02-04ER-46139 and the SC EPSCoR cost sharing program. The work at Minsk was partly supported by BRFFI under grant No. F09K-017.

**Figure captions:**

Figure 1: (color online) Energy scans (a) as a function of temperature at $\mathbf{Q} = (0.53, 0.53, 1.53)$ and (b) at $\mathbf{Q} = (0.55, 0.55, 1.55)$, which is for the transverse acoustic (TA) $[\xi\xi\xi]$ mode with $\xi = 0.47$ in (a) and $\xi = 0.45$ in (b). (a) shows the softening of the phonon close to the zone boundary with decreasing temperature from T = 320 K to 200 K. (b) Below T = 200 K the phonon peak position shows no change from T = 140 to 8 K. No anomaly of the phonon energy is observed crossing $T_c$. The solid lines are guides to the eye and the error bars are statistical in origin and represent one standard deviation.

Figure 2: (color online) (a) Superstructure peak at (1.5, 1.5, 1) (a new (0.5, 0.5, 0) type superstructure peak) is observed at temperatures below 200 K. (b) The cubic crystal structure of BKBO at room temperature. Each Bi atom is surrounded by 6 oxygen atoms forming $BiO_6$ octahedra. (c) The tetragonal crystal structure suggested by Braden *et al*.[38] at low temperature. (d) The octahedral rotation around [001] in the *ab* plane suggested by Braden *et al*. The small (black) square is the basal plane for a cubic unit cell and the large (blue) square is the basal plane for a tetragonal unit cell with lattice parameters, $a_t = \sqrt{2}a_c$ and $c_t = 2a_c$.

Figure 3: (color online) (a) and (b) [HHH] scans and integrated intensities around (0.5, 0.5, 2.5). A (0.5, 0.5, 0.5) type superstructure peak develops when the temperature is decreased. The integrated intensity in (b) shows that the structural transition is second order and the structural transition temperature is $T_s \approx 200$ K. (c) and (d) Energy scan through the elastic position of (0.5, 0.5, 1.5) peak and its integrated intensity. The data in **Q** and energy scans are least-squares fit with a Gaussian. The solid lines in (b) and (d) for the integrated intensities are guides to the eye.

Figure 4: (color online) Transverse acoustic phonon dispersion curve along [111] direction at several temperatures, showing that this phonon mode softens with decreasing temperature.

Figure 1:

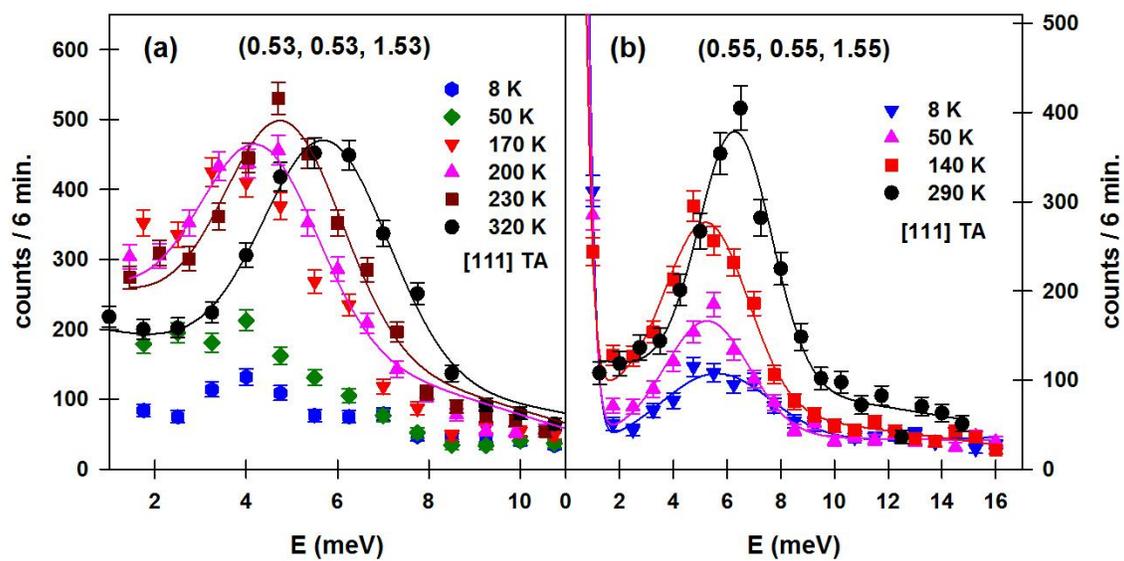

Figure 2:

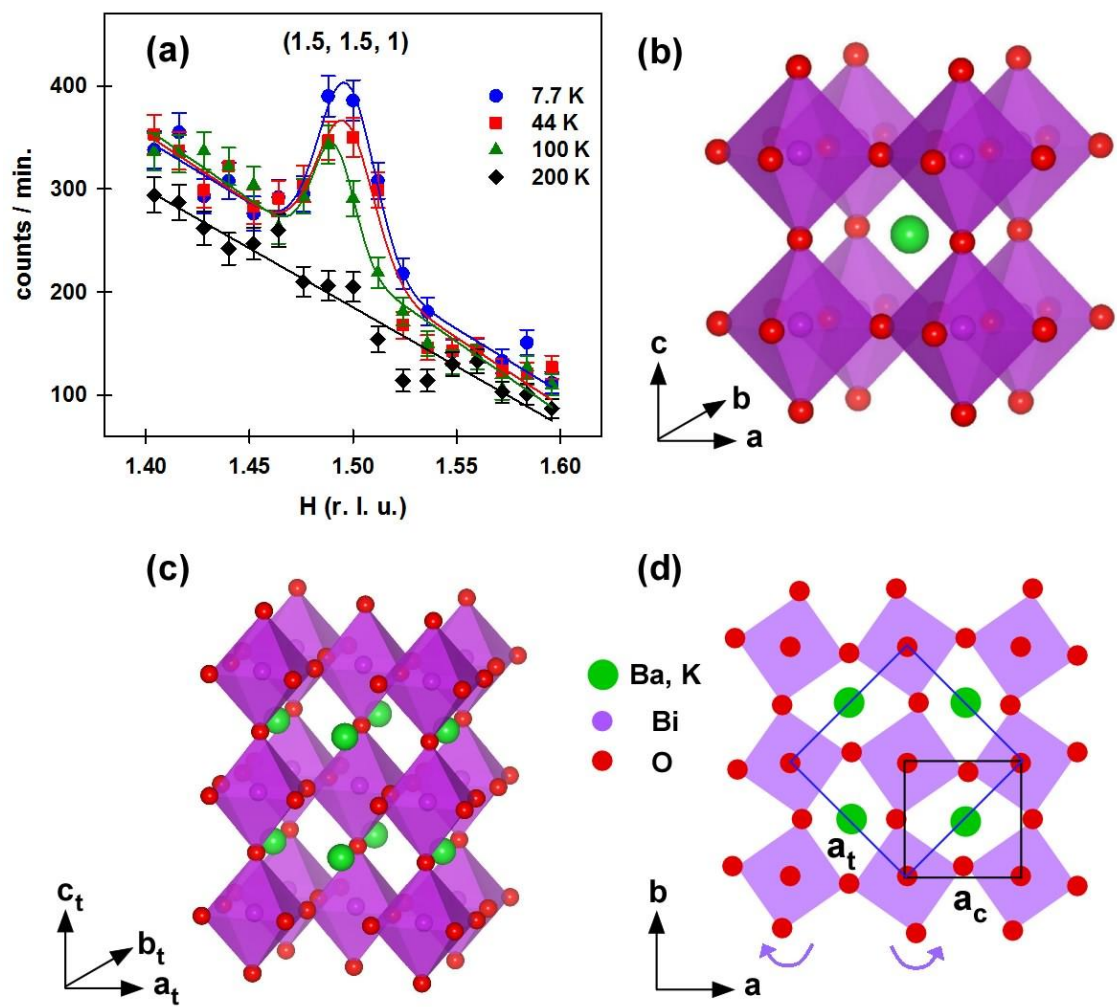

Figure 3:

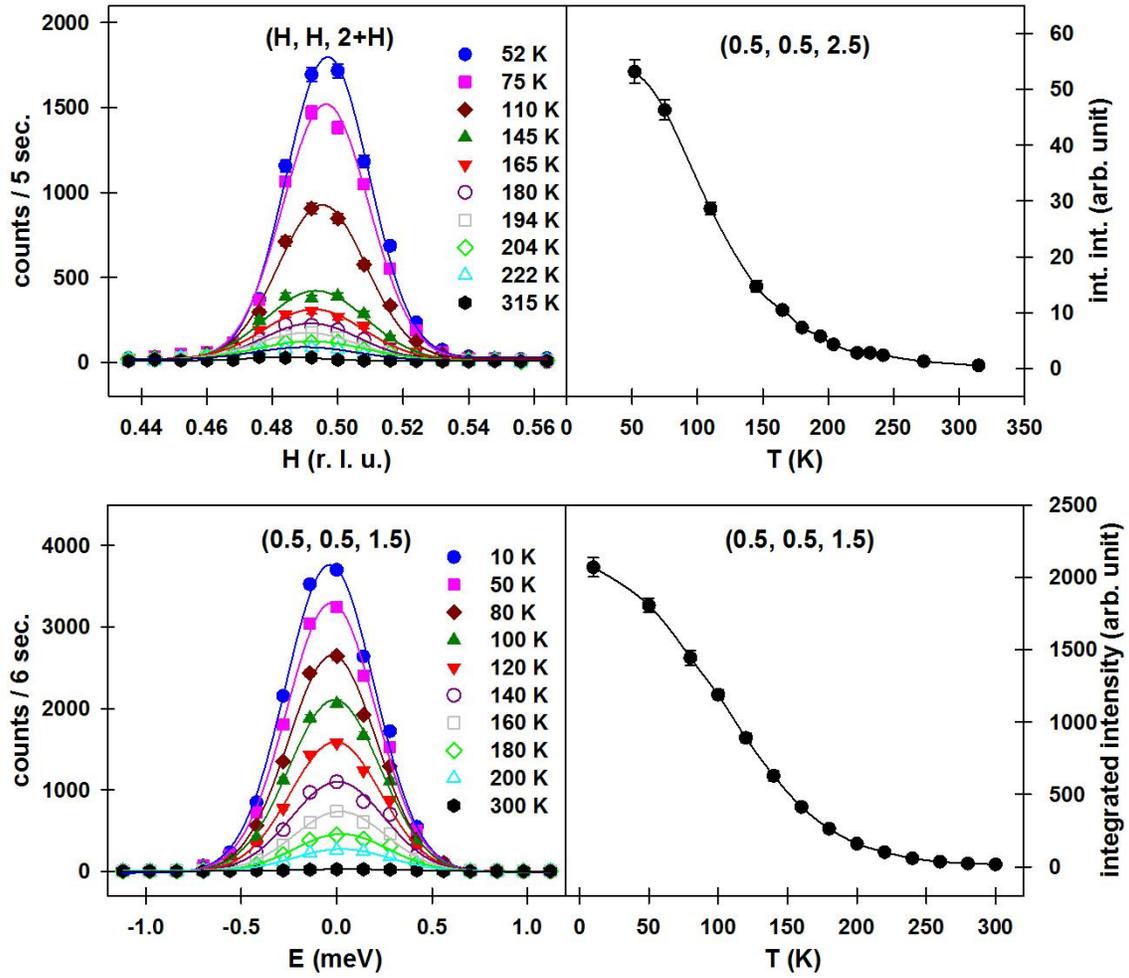

Figure 4:

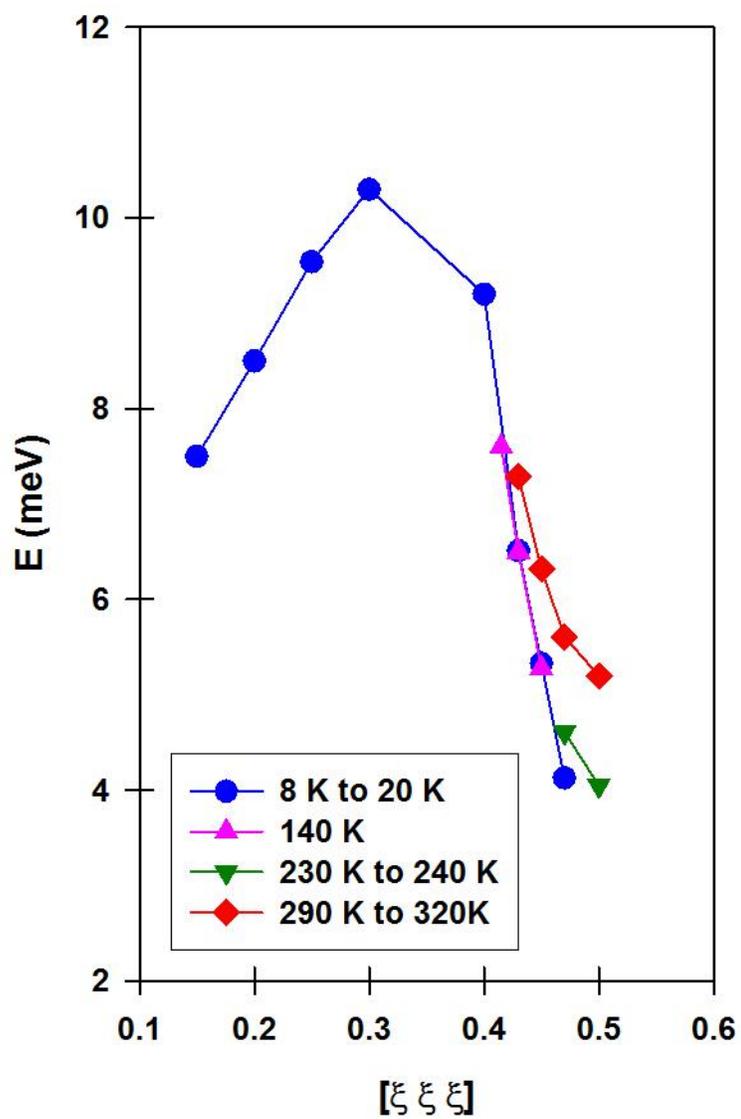